\documentclass{article}
\usepackage{spconf,amsmath,graphicx}
\usepackage{amsfonts}
\usepackage{booktabs}
\usepackage[hidelinks]{hyperref}
\usepackage[subrefformat=parens]{subcaption}
\usepackage{siunitx}
\usepackage{mathtools}
\usepackage{comment}
\usepackage{multirow}
\usepackage{makecell}

\usepackage{tikz}
\usetikzlibrary{calc}
\usetikzlibrary{tikzmark}
\usetikzlibrary{arrows, positioning,intersections,angles}
\usetikzlibrary{fit}
\usetikzlibrary{patterns}

\usepackage{xspace}
\makeatletter
\DeclareRobustCommand\onedot{\futurelet\@let@token\@onedot}
\def\@onedot{\ifx\@let@token.\else.\null\fi\xspace}

\makeatother

\def\equationautorefname~#1\null{(#1\null)}

\renewcommand{\sectionautorefname}{Section}
\renewcommand{\subsectionautorefname}{\sectionautorefname}

\let\orgautoref\autoref
\providecommand{\Autoref}[1]
{%
\def\figureautorefname{Figure}%
\def\subfigureautorefname{Figure}%
\orgautoref{#1}%
}
\renewcommand{\autoref}[1]
{%
\def\figureautorefname{Fig.}%
\def\subfigureautorefname{\figureautorefname}%
\def\sectionautorefname{Sec.}%
\def\subsectionautorefname{\sectionautorefname}%
\def\subsectionautorefname{\sectionautorefname}%
\orgautoref{#1}%
}

\newcommand{\vect}[1]{\mbox{\boldmath $#1$}}
\newcommand{\abs}[1]{\left\lvert#1\right\rvert}
\newcommand{\norm}[1]{\left\lVert#1\right\rVert}

\def\appendixautorefname~#1\null{~#1 \null}

\newcommand{\svss}{\textit{s vs. s}}
\newcommand{\svsm}{\textit{s vs. m}}
\newcommand{\mvsm}{\textit{m vs. m}}

\makeatletter
\newcommand{\figcaption}[1]{\def\@captype{figure}\caption{#1}}
\newcommand{\tblcaption}[1]{\def\@captype{table}\caption{#1}}
\makeatother

\makeatletter
\renewenvironment{description}%
   {\list{}{\leftmargin=10pt 
            \labelwidth\z@ \itemindent-\leftmargin
            }}%
   {\endlist}
\makeatother

\title{Recursive attentive pooling for extracting speaker embeddings from multi-speaker recordings}

\name{\begin{tabular}{c}Shota Horiguchi \quad Atsushi Ando \quad Takafumi Moriya \quad Takanori Ashihara \\
Hiroshi Sato \quad Naohiro Tawara \quad Marc Delcroix\end{tabular}}
\address{NTT Corporation, Japan}
\begin{document}
\ninept
\abovedisplayskip=3pt
\belowdisplayskip=3pt

\setlength\textfloatsep{10pt}
\setlength\abovecaptionskip{4pt}
\setlength\belowcaptionskip{4pt}

\maketitle
\begin{abstract}
This paper proposes a method for extracting speaker embedding for each speaker from a variable-length recording containing multiple speakers.
Speaker embeddings are crucial not only for speaker recognition but also for various multi-speaker speech applications such as speaker diarization and target-speaker speech processing.
Despite the challenges of obtaining a single speaker's speech without pre-registration in multi-speaker scenarios, most studies on speaker embedding extraction focus on extracting embeddings only from single-speaker recordings.
Some methods have been proposed for extracting speaker embeddings directly from multi-speaker recordings, but they typically require preparing a model for each possible number of speakers or involve complicated training procedures.
The proposed method computes the embeddings of multiple speakers by focusing on different parts of the frame-wise embeddings extracted from the input multi-speaker audio.
This is achieved by recursively computing attention weights for pooling the frame-wise embeddings.
Additionally, we propose using the calculated attention weights to estimate the number of speakers in the recording, which allows the same model to be applied to various numbers of speakers.
Experimental evaluations demonstrate the effectiveness of the proposed method in speaker verification and diarization tasks.
\end{abstract}

\section{Introduction}
A speaker-discriminative embedding (or \textit{speaker embedding}) is a feature extracted from an audio segment that represents the speaker's identity or voice characteristics, while excluding the content contained in the segment and other paralinguistic information.
The use of speaker embeddings has become the \textit{de facto} standard in speaker recognition tasks, i.e., speaker identification and speaker verification \cite{snyder2017deep,ji2018end,an2019deep,chung2020defence,desplanques2020ecapatdnn,zhou2021resnext}.
These tasks usually assume only a single speaker per recording, so most conventional methods cannot extract embeddings from multi-speaker recordings.
At the same time, in recent years, the use of speaker embeddings has expanded beyond speaker recognition to include speech applications for multi-speaker audio.

One example of multi-speaker applications in which speaker embeddings play an important role is speaker diarization.
For example, in a cascaded speaker diarization system, diarization is performed by dividing detected speech intervals into small segments, extracting a speaker embedding from each segment, and then clustering them \cite{park2022review}.
Since speaker embedding extraction and clustering are performed by assuming that each segment corresponds to a single speaker, the functional limitation is that overlapping speech cannot be handled.
Overlaps can be treated to some extent by post-processing \cite{diez2019bayesian,bullock2020overlap,horiguchi2021endtoend}, but embeddings extracted from overlaps are distributed in the middle of each speaker's embeddings \cite{cord2024geodesic}, which can be an obstacle to clustering.
There are end-to-end speaker diarization methods that naturally deal with overlaps such as EEND \cite{fujita2019end2,horiguchi2022encoderdecoder} and TS-VAD \cite{medennikov2020targetspeaker}.
However, they still have challenges such as the need for a large amount of high-quality simulation data for training \cite{yamashita2022improving,landini2022from,landini2023multi} and the eventual necessity of extracting speaker embeddings from multi-speaker speech segments \cite{medennikov2020targetspeaker,kinoshita2021integrating,kinoshita2021advances,he2023ansdmamse}.

Another example of a multi-speaker application using speaker embeddings is to obtain speech processing results for a speaker of interest, i.e., target speaker.
Many models for target-speaker speech processing have been proposed such as voice activity detection \cite{ding2020personal,he2021targetspeaker}, speaker extraction \cite{zmolikova2023neural,wang2018deep,wang2019voicefilter}, and speech recognition \cite{delcroix2018single,wang2020voicefilterlite,moriya2022streaming}.
Target-speaker speech processing usually assumes the availability of a single-speaker utterance of the target speaker to capture reliable speaker embeddings.
However, extracting reliable speaker embeddings from multi-speaker audio would allow us to expand the applications of target-speaker based tasks.

Despite the benefits of extracting speaker embeddings from multi-speaker audio, few studies have tackled this problem.
One possible approach is to apply speech separation in advance to obtain single-speaker recordings \cite{xiao2021microsoft}.
However, this is known to produce artifacts that may negatively affect the later processing stage \cite{sato2022learning}; that is, separation may damage the original speaker characteristics.
Another approach is to construct a model that directly extracts the embeddings of each speaker from multi-speaker audio.
However, conventional methods require preparing a model for each number of speakers \cite{han2020mirnet,cord2023teacher}, which is quite costly, and teacher-student learning is necessary to achieve reasonable performance \cite{cord2023teacher}.

In this paper, we propose a method for extracting a speaker embedding for each speaker from audio that may include multiple speakers.
Typical speaker embedding extractors have a cascaded structure consisting of i) an encoder that converts variable-length input into frame-wise embeddings and ii) a pooling module that aggregates the frame-wise embeddings into a single speaker embedding.
The proposed method enables the extraction of embeddings for each speaker, even from fully overlapped speech, by focusing on different parts of the frame-wise embeddings.
More specifically, the proposed method recursively calculates the attention weights used for pooling a variable-length sequence of frame-wise embeddings.
It is notable that this can be achieved only by adding a single linear layer to the original speaker embedding extractors.
We also propose a method that uses the calculated attention weights in estimating the number of speakers, which enables the extraction of speaker embeddings corresponding to each speaker using the same model even when the number of speakers in the input is unknown.
We demonstrate the effectiveness of the proposed method in speaker verification and diarization tasks, and also provide detailed analyses of the experimental results to clarify the behavior of the proposed method.

\section{Related work}
Various architectures have been proposed for speaker embedding extraction. 
As the encoder, previous studies have investigated architectures such as time delay neural network (TDNN) \cite{snyder2018xvectors,desplanques2020ecapatdnn}, long short-term memory \cite{wan2018generalized}, and convolutional neural network \cite{nagrani2017voxceleb,zhou2021resnext}.
For pooling methods, various alternatives to the simple temporal average pooling \cite{snyder2017deep} have been proposed, such as statistics pooling \cite{snyder2017deep}, attentive statistics pooling \cite{okabe2018attentive}, multi-head attention pooling \cite{india2019self}, vector-based attentive pooling \cite{wu2020vector}, and channel- and context-independent statistics pooling \cite{desplanques2020ecapatdnn}.
Eventually, regardless of the encoder or pooling method, the output is a single embedding, which means that a single speaker's recording is assumed as the input.

There are few studies that have investigated speaker embedding extraction from multi-speaker recordings.
MIRNet \cite{han2020mirnet} uses the attention mechanism to extract each embedding of two speakers.
It includes a processor that creates the embedding for the second speaker by swapping the channels of the first and second halves of the embedding computed for the first speaker, which fixes the number of speakers to two.
In addition, the attention weights are frame-wise ones, so in principle, it requires single-speaker segments and never works perfectly for fully overlapped speech.
In EEND-vector clustering \cite{kinoshita2021integrating,kinoshita2021advances}, multiple speakers' embeddings are extracted on the basis of estimated overlap-aware diarization results.
Here, speaker embeddings are also calculated as the weighted average of frame-wise embeddings, and thus, it cannot deal with fully overlapped speech.
Very recent work proposed a method for extracting the embedding of each speaker from a fully overlapped speech by developing a multi-headed model \cite{cord2023teacher}.
The proposal uses teacher-student learning, i.e., a model that outputs two embeddings from two-speaker audio is trained to mimic each output of a well-trained single-speaker model given single-speaker sources.
The authors reported that training the model from scratch did not work well, so using this method requires two training runs, which is time-consuming.
Also, since the model can only process two-speaker speech, it is necessary first to determine whether the input audio contains one or two speakers and then use the appropriate model.
In contrast, our approach can use the same model regardless of the number of speakers and can be trained from scratch.

\section{Method}
\begin{figure*}[t]
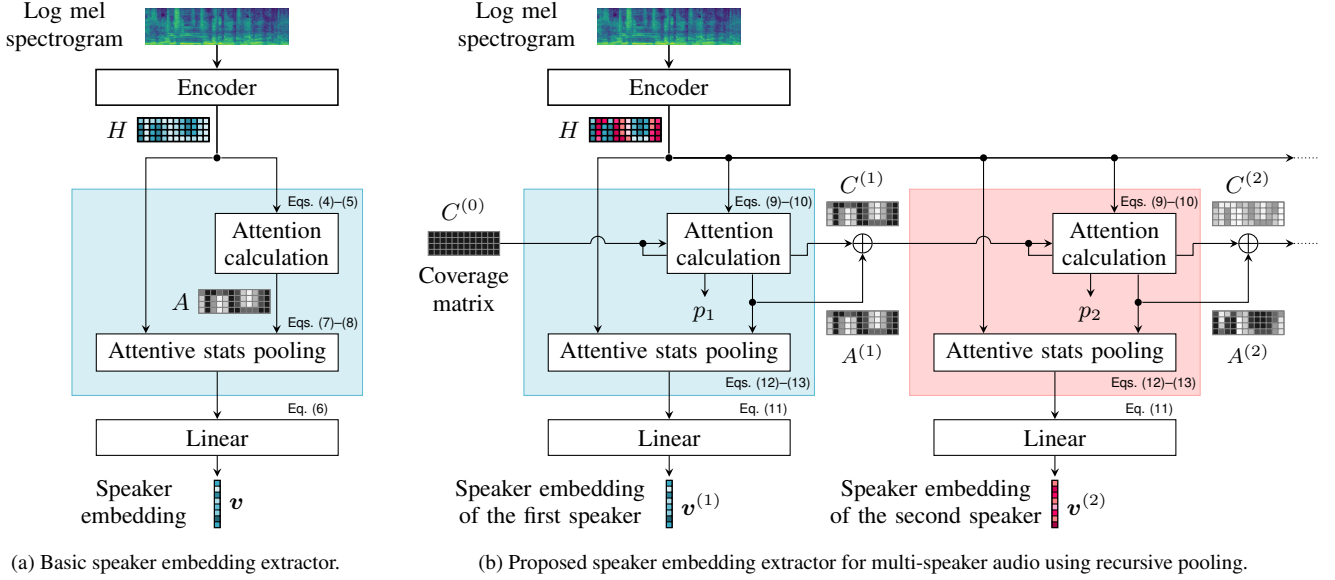

\centering
\subfloat[Basic speaker embedding extractor.]{\input{figs/diagram_conventional_eq.tex}\label{fig:diagram_conventional}}\hfill
\subfloat[Proposed speaker embedding extractor for multi-speaker audio using recursive pooling.]{\input{figs/diagram_eq.tex}\label{fig:diagram_proposed}}
\caption{Schematic diagram of the conventional and proposed methods.}
\label{fig:diagram}%
\end{figure*}

\Autoref{fig:diagram} shows a schematic diagram of a common single-speaker speaker embedding extractor (\autoref{fig:diagram_conventional}) and how we extend it to enable the decoding of each of multiple speakers' embeddings (\autoref{fig:diagram_proposed}).
Since our proposal is a method of pooling the output from the encoder, any encoder architecture that extracts a sequence of frame-wise embeddings from input recording can be adopted.

\subsection{Review of channel- and context-dependent attentive statistics pooling}
We first review the channel- and context-dependent attentive statistics pooling \cite{desplanques2020ecapatdnn} on which our proposed method is based.
This technique is widely used in the modern speaker embedding extractors \cite{wang2021maccif,mun2023frequency,zhao2023pcf}.
It performs weighted averaging of frame-wise embeddings using attention over the time axis for each dimension to maintain information meaningful in the computation of speaker embedding.
In this respect, it is similar to mask-based speech enhancement, which extracts only those time-frequency bins that contain speech.

Given a $T$-length sequence of $D$-dimensional frame-wise embeddings $H\coloneqq[\vect{h}_1,\dots,\vect{h}_T]\in\mathbb{R}^{D\times T}$ from the encoder, each is first expanded with their mean and standard deviation vectors $\vect{\mu}$ and $\vect{\sigma}$ to consider the global context as
\begin{align}
    \vect{e}_t&=\vect{h}_t\oplus\vect{\mu}\oplus\vect{\sigma}\in\mathbb{R}^{3D},\\
    \vect{\mu}&=\frac{1}{T}\sum_{\tau=1}^T\vect{h}_\tau\in\mathbb{R}^D,\\ \vect{\sigma}&=\sqrt{\frac{1}{T}\sum_{\tau=1}^{T}\vect{h}_{\tau}\odot\vect{h}_{\tau}-\vect{\mu}\odot\vect{\mu}}\in\mathbb{R}^D.
\end{align}
where $\oplus$ denotes vector concatenation and $\odot$ denotes the Hadamard product.
The attention weights $A\coloneqq\left[\vect{a}_1,\dots,\vect{a}_T\right]$ to aggregate the expanded embeddings $\vect{e}_t$ is calculated by
\begin{align}
    A&=\mathsf{Softmax}\left(\left[\tilde{\vect{a}}_1,\dots,\tilde{\vect{a}}_T\right]\right)\in\left(0,1\right)^{D\times T},\label{eq:attn}\\
   \tilde{\vect{a}}_t&=W_2 f\left(W_1\vect{e}_t+\vect{b}_1\right)+\vect{b}_2\in\mathbb{R}^D,\label{eq:attn_logit}
\end{align}
where $\mathsf{Softmax}\left(\cdot\right)$ denotes the row-wise softmax function, $W_1\in\mathbb{R}^{D'\times 3D}$ and $\vect{b}_1\in\mathbb{R}^{D'}$ denotes the weight and bias of the first linear layer, $W_2\in\mathbb{R}^{D\times D'}$ and $\vect{b}_2\in\mathbb{R}^{D}$ are those of the second layer, and $f\left(\cdot\right)$ is the rectified linear unit, respectively.
Speaker embedding $\vect{v}$ is then computed as
\begin{equation}
    \vect{v}=W_o\left(\tilde{\vect{\mu}}\oplus\tilde{\vect{\sigma}}\right)+\vect{b}_o\in\mathbb{R}^{E},\label{eq:embed}
\end{equation}
where $W_o\in\mathbb{R}^{E\times 2D}$ and $\vect{b}_o\in\mathbb{R}^E$ are the parameters of the last linear layer to obtain $E$-dimensional speaker embeddings.
$\tilde{\vect{\mu}}$ and $\tilde{\vect{\sigma}}$ are the weighted mean and standard deviation vectors, respectively, calculated using the attention weights as:
\begin{align}
    \tilde{\vect{\mu}}&=\sum_{\tau=1}^{T}\vect{a}_{\tau}\odot\vect{h}_{\tau}\in\mathbb{R}^D,\label{eq:embed_mean}\\
    \tilde{\vect{\sigma}}&=\sqrt{\sum_{\tau=1}^{T}\vect{a}_{\tau}\odot\vect{h}_{\tau}\odot\vect{h}_{\tau}-\tilde{\vect{\mu}}\odot\tilde{\vect{\mu}}}\in\mathbb{R}^D.\label{eq:embed_std}
\end{align}

\subsection{Recursive embedding extraction and speaker counting}\label{sec:proposed}
Since conventional pooling methods aggregate a variable-length sequence of frame-wise embeddings into a single embedding, single-speaker audio is assumed as input.
In this paper, we extend the channel- and context-dependent attentive statistics pooling to extract as many speaker embeddings as there are speakers in the input audio.
If frame-wise embeddings that can assume speaker sparsity are achieved, we can use the mask-based speech separation approach to extract an embedding for each speaker.
In particular, by calculating masks recursively as in recurrent selective attention network (RSAN) \cite{kinoshita2018listening}, it is possible for a single model to extract embeddings for an arbitrary number of speakers.
RSAN uses the principle that the sum of masks for each source is constant, but it does not hold in the case of attention weights in the speaker embedding extractor.
Therefore, we instead introduce the coverage mechanism \cite{tu2016modeling,see2017get}, which was originally proposed for neural machine translation to prevent over- or under-translation, to monitor which parts of the frame-wise embeddings have already been used for decoding speaker embeddings.
Note that we provide a general formulation of the proposed method for an arbitrary number of speakers in this section, while we assume in the experiments that the input is either one- or two-speaker audio.
This is because we can easily remove regions where there are no speakers using voice activity detection, and in practice, there are often not more than two speakers speaking at the same time \cite{chen2020continuous}.

In the proposed method, attention weights for the $n$-th speaker are calculated from not only the frame-wise embeddings but also the cumulative sum of the previously calculated attention weights.
Instead of \autoref{eq:attn} and \autoref{eq:attn_logit}, we calculate the attention weights for each speaker $A^{(n)}\coloneqq\left[\vect{a}_1^{(n)},\dots,\vect{a}_T^{(n)}\right]$ that appear in the input audio in a recursive manner as follows:
\begin{align}
    A^{(n)}&=\mathsf{Softmax}\left(\tilde{\vect{a}}_1^{(n)}\dots,\tilde{\vect{a}}_T^{(n)}\right)\in\left(0,1\right)^{D\times T},\label{eq:proposed_attn}\\
   \tilde{\vect{a}}_t^{(n)}&=W_2 f\left(W_1\vect{e}_t+\vect{b}_1+W_c\vect{c}_t^{(n)}\right)+\vect{b}_2\in\mathbb{R}^D.\label{eq:proposed_attn_logit}
\end{align}
Here, $C^{(n)}=\left[\vect{c}_1^{(n)},\dots,\vect{c}_T^{(n)}\right]$ where $\vect{c}_t^{(n)}=\sum_{r=0}^{n-1} \vect{a}_t^{(r)}\in\mathbb{R}_{\geq0}^{D}$ is the coverage matrix defined as the cumulative sum of the previous attentions, and $W_c\in\mathbb{R}^{D'\times D}$ is the only additional learnable parameter needed in the proposed method.
Note that $\vect{a}_t^{(0)}$ is defined as a $D$-dimensional zero vector.
Here, we aim to extract multiple embeddings for different speakers from the same input audio by applying attention to regions that have not been focused on by the $(n-1)$-th speaker when computing the embedding for the $n$-th speaker.
This requires speaker sparsity in the frame-wise embeddings from the encoder, similar to the mask-based speech separation.
The proposed method achieves this by end-to-end optimization of the entire model.

Finally, the embedding of the $n$-th speakers is extracted as the same manner in \autoref{eq:embed}--\autoref{eq:embed_std} as follows:
\begin{align}
    \vect{v}^{(n)}&=W_o\left(\tilde{\vect{\mu}}^{(n)}\oplus\tilde{\vect{\sigma}}^{(n)}\right)+\vect{b}_o\in\mathbb{R}^{E},\label{eq:embed_proposed}\\
    \tilde{\vect{\mu}}^{(n)}&=\sum_{\tau=1}^{T}\vect{a}_{\tau}^{(n)}\odot\vect{h}_{\tau}\in\mathbb{R}^D,\label{eq:embed_mean_proposed}\\
    \tilde{\vect{\sigma}}^{(n)}&=\sqrt{\sum_{\tau=1}^{T}\vect{a}_{\tau}^{(n)}\odot\vect{h}_{\tau}\odot\vect{h}_{\tau}-\tilde{\vect{\mu}}^{(n)}\odot\tilde{\vect{\mu}}^{(n)}}\in\mathbb{R}^D.\label{eq:embed_std_proposed}
\end{align}

To decide when to stop the recursive attention generation process, we monitor whether any parts of the frame-wise embeddings still attract attention.
Since attention is normalized along the time axis using the softmax function (unlike mask-based source separation), we use the values before normalization $\tilde{\vect{a}}_t^{(n)}$ for the monitoring.
Speaker existence probability $p_n$, which indicates whether the $n$-th speaker really exists in the input audio, is estimated as
\begin{equation}
    p_n=\frac{1}{1+\exp\left(-\frac{1}{T}\sum_{t=1}^T\vect{w}\cdot\tilde{\vect{a}}_t^{(n)}+b\right)},
\end{equation}
where $\vect{w}\in\mathbb{R}^D$ and $b\in\mathbb{R}$ are the learnable parameters.
During inference, thresholding $p_n$ acts as a stopping condition, where the threshold value is set to $0.5$ in this paper.

\subsection{Training objective}
We assume that the model outputs $N$ speaker embeddings $\mathcal{V}\coloneqq\left\{\vect{v}_1,\dots,\vect{v}_N\right\}$, each of which corresponds to one of the individuals in the training speaker set $\mathcal{Y}\coloneqq\left\{y_1,\dots,y_S\right\}$.
The model also output existence probabilities $p_1,\dots,p_{N+1}$ for the $N+1$ speakers.
The training objective to be minimized is determined as follows:
\begin{equation}
    \mathcal{L}=\mathcal{L}_\text{spk}+\alpha\mathcal{L}_\text{cnt},\label{eq:loss}
\end{equation}
where $\alpha$ is a weight factor that is fixed to 0.1 in this paper.

The first term, $\mathcal{L}_\text{spk}$, in \autoref{eq:loss} aims to optimize the similarity metric of speaker embeddings.
We train the network as a multi-class classifier using the permutation-free loss
\begin{equation}
    \mathcal{L}_\text{spk}=\min_{\left(y_1^\phi,\dots,y_N^\phi\right)\in\Phi\left(\mathcal{Y'}\right)}\frac{1}{N}\sum_{n=1} ^{N}\ell\left(y_n^\phi, \vect{v}^{(n)}\right),
\end{equation}
where $\mathcal{Y'}\subseteq\mathcal{Y}$ is the set of speakers in the input audio and $\Phi\left(\mathcal{Y'}\right)$ is all permutation of $\mathcal{Y'}$.
$\ell\left(y,\vect{v}\right)$ is the additive angular margin (AAM) softmax loss \cite{deng2022arcface} defined as
\begin{align}
    \ell\left(y,\vect{v}\right)&=-\log\frac{e^{s\cos\left(\theta_y+m\right)}}{e^{s\cos\left(\theta_y+m\right)}+\sum_{n=1,n\neq y}^{N}e^{s\cos\theta_n}},\\
    \theta_y&=\arccos\frac{\vect{w}_y\cdot\vect{v}}{\norm{\vect{w}_y}\norm{\vect{v}}},
\end{align}
where $m>0$ denotes the margin, $s>0$ is the scaling factor, and $\vect{w}_y\in\mathbb{R}^{E}$ is the learnable proxy of identity $y$.

The second term, $\mathcal{L}_{\text{cnt}}$, in \autoref{eq:loss} aims to optimize the speaker counting accuracy, which is defined from cross-entropy as
\begin{equation}
    \mathcal{L}_{\text{cnt}}=-\frac{1}{N+1}\left(\sum_{n=1}^N\log\left(1-p_n\right)+\log p_{N+1}\right).
    \label{eq:l_cnt_general}
\end{equation}
Here, $p_n$ for $1\leq n \leq N$ is optimized to be one, and $p_{N+1}$ is optimized to be zero.
As mentioned in \autoref{sec:proposed}, this paper assumes that the input contains one or two speakers, so we use the simplified alternative below to \autoref{eq:l_cnt_general} only to check if the second speaker exists:
\begin{equation}
    \mathcal{L}_{\text{cnt}}=\begin{cases}
        -\log p_2 & (N=1),\\
        -\log\left(1-p_2\right) & (N=2).
    \end{cases}
\end{equation}

\subsection{Inference-time length mismatch correction}
It is common to align the length of each sample in a single minibatch during training for efficient batch processing.
However, models are usually required to process variable-length inputs during inference.
The conventional method does not suffer from this mismatch because $\tilde{\vect{a}}_t$ in \autoref{eq:attn_logit} is calculated independently for each frame.
On the other hand, the proposed method includes $\vect{c}_t^{(n)}$ in \autoref{eq:proposed_attn_logit}, which depends on the sequence length for the second and subsequent speakers because $\vect{a}_t^{(n)}$ is computed with the softmax over the time dimension as seen in \autoref{eq:proposed_attn}.
We use the following equation instead of \autoref{eq:proposed_attn_logit} only during inference so that any mismatch in audio length between training and inference does not affect the results:
\begin{equation}
    \tilde{\vect{a}}_t^{(n)}=W_2 f\left(W_1\vect{e}_t+\vect{b}_1+\frac{T_\text{infer}}{T_\text{train}}W_c\vect{c}_t^{(n)}\right)+\vect{b}_2,\label{eq:bias_corrected}
\end{equation}
where $T_\text{train}$ is the sequence length of frame-wise embeddings fixed during training and $T_\text{infer}$ is their length extracted from inference speech.
This improves inference performance for the second and subsequent speakers by making the expected value of attention weights per frame the same during training and inference, which eliminates the effect of sequence-length mismatch.

\section{Experimental settings}
\subsection{Training}
We used the concatenation of the VoxCeleb1 dev set and VoxCeleb2 dev set \cite{nagrani2020voxceleb} for training; the result consisted of 1,240,651 utterances from 7,205 speakers.
The VoxCeleb2 test set was used for validation.

As the network architecture, we tested three different encoders to show the generality of the proposed method: x-vector encoder consisting of a five-stacked TDNN ($D=1500$) \cite{snyder2018xvectors}, ResNet34 ($D=2560$) \cite{he2016deep,wang2020investigation,cord2023teacher}, and ECAPA-TDNN with 1024 channels ($D=1536$) \cite{desplanques2020ecapatdnn}.
Channel- and context-aware attentive statistics pooling and the proposed method were used as the pooling methods regardless of encoder type.
The output dimension of the final linear layer was set to $E=192$.
The input to each network was a sequence of mean-normalized 80-dimensional log mel filterbanks extracted with a window length of \SI{25}{\ms} and shift of \SI{10}{\ms}.
This yielded 100 frame-wise embeddings per second for x-vector and ECAPA-TDNN, and 12.5 for ResNet34.

During training, each utterance in a mini-batch was \SI{3}{\second} in duration and augmented using noise \cite{snyder2015musan} with a probability of 0.5.
It was further reverberated using simulated room impulse responses \cite{ko2017study} with a probability of 0.5.
The mini-batch size was 256 for the single-speaker baselines and 384 (256 single-speaker audio and 128 two-speaker audio) for the proposed method.
Each two-speaker audio was a fully overlapped mixture generated on the fly during training.
Signal-to-interference ratio (SIR) used to generate mixtures lay in a range of $\left[-5, 5\right] \si{dB}$, following the separation benchmark~\cite{hershey2016deep}.

Each model was trained for 80 epochs using the Adam optimizer \cite{kingma2015adam} with a cyclical learning rate.
Each cycle consisted of 20 epochs, with the first 1,000 iterations taken as warm-up, followed by cosine annealing for the remaining iterations.
The peak learning rate at the first cycle was set to 0.001 and 0.0005 for the baselines and proposed method, respectively, and was decayed by a factor of 0.75 with each cycle.
The margin and scaling factor of AAM-softmax were set to 0.2 and 30, respectively.

\subsection{Evaluation}
\subsubsection{Speaker verification}
Following the previous study \cite{cord2023teacher}, speaker verification performance was evaluated under the three scenarios detailed below:
\begin{description}
    \item[\svss:] This is the standard single-speaker audio vs. single-speaker audio scenario. The evaluation used the VoxCeleb1 test set, a.k.a. VoxCeleb1-O, consisting of 37,611 trials from 40 speakers.
    \item[\svsm:] This is the single-speaker audio vs. two-speaker audio scenario.
    Each positive sample in \svsm{} is a pair in which the speaker of the single-speaker audio is one of the speakers in the mixture, while a negative sample is a pair in which the speaker of the single-speaker audio is not included in the mixture.
    We used the same evaluation set used in the conventional study \cite{cord2023teacher}, i.e., 37,611 trials extended from VoxCeleb1-O by mixing interference speech with random SIR.
    \item[\mvsm:] This is the two-speaker audio vs. two-speaker audio scenario.
    A positive (negative) sample in \mvsm{} is a pair in which one speaker in one of the mixture is included (not included) in the other mixture.
    Note that pairs of mixtures in which both speakers were the same were not included.
    In this scenario, two evaluation protocols were used: \textit{any spk} and \textit{per spk}. In the \textit{any spk} protocol, only the largest similarity value among all embeddings combinations extracted from each audio was used for evaluation, whereas in the \textit{per spk} scenario, the other pair was also used as a negative pair.
    Also note that the \textit{per spk} protocol assumes that two embeddings are extracted from any mixture, so it can be evaluated only if the model is capable of outputting two embeddings and the oracle number of speakers is given.
    Here too, we used the evaluation set of 37,611 trials used in the conventional study \cite{cord2023teacher}.
\end{description}

As the evaluation metrics, equal error rate (EER) and minimum detection cost function (minDCF) were used, where the prior probability for minDCF was set to 0.01 for \svss{} and 0.05 for \svsm{} and \mvsm{} as in the conventional study~\cite{cord2023teacher}.
We used cosine similarity for scoring each trial.

\subsubsection{Speaker diarization}
Speaker diarization performance was evaluated using the LibriCSS dataset \cite{chen2020continuous} and AMI Mix-Headset corpus \cite{carletta2007unleashing}.
We used auto-tuning spectral clustering (SC) \cite{park2020auto} and its extension to deal with overlapping speech (SC-OL) \cite{raj2021multiclass} as the baseline methods.
Speech and overlapped segments were given by the oracle voice activity detector and overlap detector.
Speaker embeddings for clustering were extracted with \SI{1.5}{\second} window with \SI{0.75}{\second} shift using the ECAPA-TDNN-based model.
For the proposed method, we extracted one speaker embedding from single-speaker segments and two speaker embeddings from overlapped segments, and applied SC with cannot-link constraints such that a pair of embeddings from the same segments were never assigned to the same cluster.
We used diarization error rate (DER) without collar tolerance as the evaluation metric.

\begin{table*}
    \caption{Single- and multi-speaker verification performance evaluated using EERs (\%) and minDCF.}\label{tbl:main_results}
    \centering
    \sisetup{detect-weight,mode=text}
    \resizebox{\linewidth}{!}{%
    \begin{tabular}{@{}llccS[table-format=1.2]S[table-format=1.2]*{3}{S[table-format=2.2]S[table-format=1.2]}@{}}
        \toprule
        &&&\multicolumn{1}{c}{\multirow{2.92}{*}{\makecell{Speaker\\counting}}}&\multicolumn{2}{c}{\svss}&\multicolumn{2}{c}{\svsm}&\multicolumn{2}{c}{\mvsm{} \textit{(any spk)}}&\multicolumn{2}{c@{}}{\mvsm{} \textit{(per spk)}}\\\cmidrule(lr){5-6}\cmidrule(lr){7-8}\cmidrule(lr){9-10}\cmidrule(l){11-12}
        ID&Encoder&\#Output&&{EER}&{minDCF}&{EER}&{minDCF}&{EER}&{minDCF}&{EER}&{minDCF}\\\midrule
        \multicolumn{5}{@{}l}{\textbf{Results from the reference papers}}\\
        \texttt{R1}&x-vector \cite{jung2024espnet}& 1&-&1.81 &0.13&{-}&{-}&{-}&{-}&{-}&{-}\\
        \texttt{R2}&ECAPA-TDNN \cite{desplanques2020ecapatdnn} &1&-& 0.87&0.11&{-}&{-}&{-}&{-}&{-}&{-}\\
        \texttt{R3}&ResNet34 (teacher) \cite{cord2023teacher} &1&-& 1.06 & 0.16 & 18.2&0.57&47.6&1.00&{-}&{-}\\
        \texttt{R3'}&\ \ + ResNet34 (student) \cite{cord2023teacher} &1 or 2& Oracle&\multicolumn{2}{c}{(Same to \texttt{R3})}& 9.1 & 0.46 & 15.3&0.74&14.1&0.74\\\midrule
        \multicolumn{5}{@{}l}{\textbf{Results based on our implementation}}\\
        \texttt{S1}&x-vector &1&-& 1.65 & 0.16 & 20.72 & 0.61 & 31.48 & 0.87 & {-} & {-}\\
        \texttt{S2}&\ \ + Proposed method& 1 or 2& Estimated&1.83&0.17&9.16&0.37&16.85&0.61&{-}&{-}\\
        \texttt{S2'}&\ \ \ \ + Oracle \# of speakers &1 or 2&Oracle& 1.82 & 0.17 & 8.16 & 0.37 & 15.11 & 0.62 & 10.63 & 0.54\\
        \texttt{S3}&ResNet34 &1&-& 1.09 & 0.11 & 20.85 & 0.57 & 32.01 & 0.83&{-}&{-}\\
        \texttt{S4}&\ \ + Proposed method& 1 or 2&Estimated&1.20&0.12&7.83&0.33&15.65&0.60&{-}&{-}\\
        \texttt{S4'}&\ \ \ \ + Oracle \# of speakers &1 or 2&Oracle&1.19&0.12&7.47&0.33&15.04&0.61&12.15&0.59\\
        \texttt{S5}&ECAPA-TDNN &1&-& 0.88 &0.09 & 24.51 & 0.59 & 35.26 & 0.84&{-}&{-}\\
        \texttt{S6}&\ \ + Proposed method& 1 or 2& Estimated&1.20&0.12&7.71&0.29&14.13&0.50&{-}&{-}\\
        \texttt{S6'}&\ \ \ \ + Oracle \# of speakers &1 or 2& Oracle& 1.17&0.12&6.35&0.28&11.97&0.50&8.34&0.41\\
        \bottomrule
    \end{tabular}%
    }
\end{table*}

\section{Results}
\subsection{Speaker verification}
\subsubsection{Main results}
The experimental results from speaker verification are shown in \autoref{tbl:main_results}.
The first four rows (\texttt{R1}--\texttt{R3'}) show the values of the conventional methods reported in the cited papers \cite{jung2024espnet,desplanques2020ecapatdnn,cord2023teacher}.
If we compare the \svss{} results of \texttt{R1}--\texttt{R3} with our reimplemented systems (\texttt{S1}, \texttt{S3}, \texttt{S5}), we can safely say that they have been mostly reproduced, while some mismatch in training strategies, such as data augmentation and the number of training iterations, yielded slight differences.
However, since these systems extract only one speaker embedding from input audio, the verification performance is quite poor in the \svsm{} and \mvsm{} scenarios as they involve multi-speaker audio.
The proposed method (\texttt{S2}, \texttt{S4}, \texttt{S6}) significantly improved the results for \svsm{} and \mvsm{} with small degradation in \svss.
Given the oracle number of speakers (\texttt{S2'}, \texttt{S4'}, \texttt{S6'}), the speaker verification performance is further improved.
Even though the proposed method uses the same model regardless of the number of speakers, it significantly outperformed the conventional method that used different models for each number of speakers \cite{cord2023teacher}.

Comparing the encoder types, even though ResNet34 (\texttt{S4'}) and ECAPA-TDNN (\texttt{S6'}) had similar verification performance on \svss, ResNet34 performed worse in the multi-speaker scenarios, especially when \mvsm, and was less accurate than as x-vector (\texttt{S2'}).
One possible reason for this is that the encoder output of the x-vector or ECAPA-TDNN retains the sequence length of the input log mel spectrogram, whereas that of ResNet34 has lower temporal resolution due to compressing the sequence length by one-eighth through the convolutional layers, and thus speaker sparsity cannot be assumed for each dimension of the embedding sequence output from the encoder.
The following sections provide detailed analyses of the models based on ECAPA-TDNN (\texttt{S5,S6,S6'}).

\subsubsection{Variable length evaluation}
\begin{table}[t]
    \centering
    \sisetup{detect-weight,mode=text}
    \caption{EERs for various durations without inference-time length mismatch correction \autoref{eq:proposed_attn_logit} and with correction \autoref{eq:bias_corrected} using \texttt{S6'}.}\label{tbl:variable_length}
    \resizebox{\linewidth}{!}{%
    \begin{tabular}{@{}l*{7}{S[table-format=2.2]}@{}}
    \toprule
    &\multicolumn{1}{c}{\svss}&\multicolumn{2}{c}{\svsm}&\multicolumn{2}{c}{\mvsm{} \textit{(any spk)}}&\multicolumn{2}{c@{}}{\mvsm{} \textit{(per spk)}}\\\cmidrule(lr){3-4}\cmidrule(lr){5-6}\cmidrule(l){7-8}
    Duration&&{\autoref{eq:proposed_attn_logit}} &{\autoref{eq:bias_corrected}}&{\autoref{eq:proposed_attn_logit}} &{\autoref{eq:bias_corrected}}&{\autoref{eq:proposed_attn_logit}} &{\autoref{eq:bias_corrected}}\\\midrule
    \SI{1}{\second}&16.41&28.02&27.44&36.52&35.64&23.84&23.54\\
    \SI{2}{\second}&5.18&14.70&14.61&23.43&23.14&15.96&15.83\\
    \SI{3}{\second} (matched)&2.67&9.77&9.77&17.31&17.31&11.92&11.92\\
    \SI{5}{\second}&1.39&6.95&6.84&12.76&12.61&9.09&8.90\\
    \SI{10}{\second}&1.20&7.07&6.33&13.34&11.97&9.83&8.35\\
    Original&1.17&7.40&6.35&13.84&11.97&10.30&8.34\\
    \bottomrule
    \end{tabular}%
    }
\end{table}

\autoref{tbl:variable_length} shows the EERs when varying the lengths of input utterances using \texttt{S6'}.
For the $t$-second audio evaluation, the score was calculated using only the first $t$ seconds of each trial pair.
Again, the audio length during training was \SI{3}{\second}.
As in the literature \cite{jung2019short}, short-duration utterances significantly degraded the EERs of the \svss{} scenario, while no disadvantages due to long utterance lengths were seen.
However, in the multi-speaker scenarios, i.e., \svsm{} and \mvsm{}, degradation of EERs was observed even with increasing utterance lengths when \autoref{eq:proposed_attn_logit} was used for attention weight calculation.
The results clearly show that introducing inference-time length mismatch correction in \autoref{eq:bias_corrected} improved the EERs in all cases where there is a mismatch in utterance length.
In addition, the disadvantage of using longer utterances than used in training almost disappeared.

\subsubsection{Detailed analyses}
This section provides detailed analyses of the proposed method including speaker counting, the effect of SIR, and attention weights.
For the analyses, we used the \svss{} trials and the positive pairs from the \svsm{} trials.
Each positive example in the \svsm{} trials consisted of a pair of single-speaker and two-speaker audio, where one speaker of the two-speaker audio is identical to the speaker of the single-speaker audio.
For the purpose of analyses, this speaker is considered as the target speaker and the other as the interference speaker to calculate SIR $r_\text{SIR}$.
For simplicity, we denote the positive case of \svss{} by $r_\text{SIR}=\infty$ because it contains only the sounds of the target speaker, and the negative case by $r_\text{SIR}=-\infty$ because it contains only the sounds of the inference speaker.

\begin{table}
    \centering
    \sisetup{detect-weight,mode=text}
    \renewrobustcmd{\bfseries}{\fontseries{b}\selectfont}
    \renewrobustcmd{\boldmath}{}
    \newrobustcmd{\B}{\bfseries}
    \caption{SIR-wise speaker counting accuracy (\%) of $\texttt{S6}$. The results corresponding to the correct prediction are \textbf{bolded}.}\label{tbl:speaker_counting}
    \resizebox{\linewidth}{!}{%
    \begin{tabular}{@{}lS[table-format=3.1]*{5}{S[table-format=2.1]}@{}}
    \toprule
    &\svss{}&\multicolumn{5}{c@{}}{\svsm{}}\\\cmidrule(lr){2-2}\cmidrule(l){3-7}
    $\abs{r_\text{SIR}}$ (\si{\dB})&${\infty}$&{$\left[0,5\right)$}&{$\left[5,10\right)$}&{$\left[10,15\right)$}&{$\geq15$}&{All}\\\midrule
    Predicted as 1 speaker&\B 100.0 &0.1&2.6&33.0&90.6&10.6\\
    Predicted as 2 speakers&0.0&\B 99.9&\B 97.4&\B 67.0 &\B 9.4&\B 89.4\\
    \midrule
    \end{tabular}%
    }
\end{table}
\autoref{tbl:speaker_counting} shows the accuracy of speaker counting for each SIR range.
Single-speaker recordings were rarely estimated as having two speakers, but the speaker counting accuracies of two-speaker recordings depend on the absolute SIR.
When the absolute SIR was below \SI{10}{\dB}, two-speaker estimates were accurate.
On the other hand, in the extreme case of SIR ($>$\SI{15}{\dB}), a single-speaker was estimated with a probability of more than \SI{90}{\percent}.

\begin{figure}
    \includegraphics[width=\linewidth]{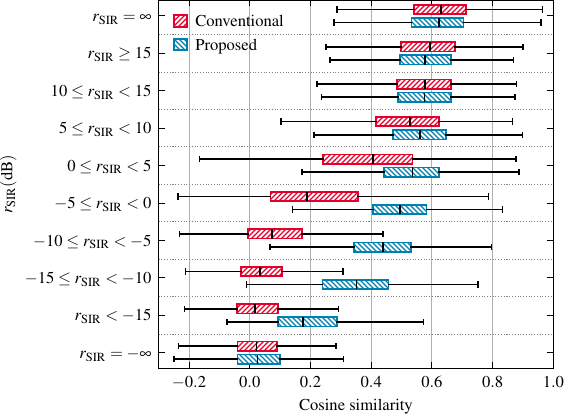}
    \caption{SIR-wise cosine similarity scores.}\label{fig:similarity}
\end{figure}

\Autoref{fig:similarity} shows the relationship between SIR and the (maximum) cosine similarity between embeddings extracted from a pair of recordings.
Note that the oracle number of speakers was given in this case.
When only one embedding was extracted from audio as in the conventional methods (red boxes in \autoref{fig:similarity}), the interference speaker became more dominant as SIR became smaller and the cosine similarity decreased.
Using the proposed method (blue boxes in \autoref{fig:similarity}), it was possible to extract embedding for each speaker, which kept high cosine similarity ($>0.4$) even when the SIR decreased up to \SI{-10}{\dB}.

\begin{figure}[t]
    \centering
    \includegraphics[width=\linewidth]{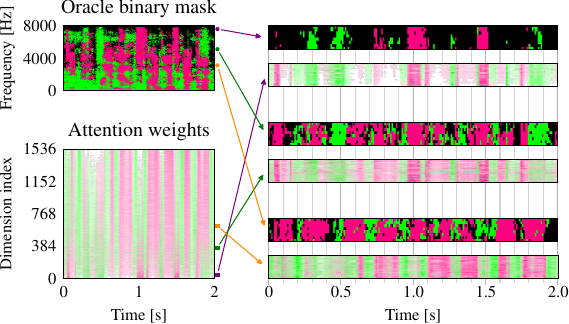}
    \caption{Visualization of attentions calculated for two speakers in a single mixture (bottom left) compared with the oracle binary mask (top left). The patterns commonly seen in the two are also cropped and shown enlarged (right).}\label{fig:attention}
\end{figure}

Finally, we visualized the ideal binary mask and computed attention weights of the example mixtures from the same and different gender pairs in \autoref{fig:attention}.\footnote{Note that the order of dimensions of the attention weights was rearranged for visualization purposes, which is valid because the weights of the speaker extractor are permutation-free.}
First, it is observed that most of the time-dimension bins had at most one speaker with high attention weight since one of the colors is dominant in the visualized attention weights (\autoref{fig:attention} left bottom).
This indicates the sparseness of speakers in the embedding sequence, similar to the sparseness of signals in the time-frequency domain.
When focusing on each specific frame, in some frames, it can be seen that different speakers attracted attention depending on the dimension.
For example, focusing on \SIrange{1.8}{1.9}{\second} in \autoref{fig:attention} right, the top and middle figures show the attention in green, whereas the bottom figure shows the attention in pink.
This indicates that, in contrast to the conventional method \cite{han2020mirnet}, the model can extract information from the same frame for different speakers depending on the dimension, similar to the oracle binary mask.
Next, when focusing on each specific dimension, the extracted attention exhibited the same patterns as the oracle binary masks (\autoref{fig:attention} right).
This indicates that the model was able to internally acquire functions similar to source separation.

\subsection{Speaker diarization}
\begin{table}[t]
    \centering
    \sisetup{detect-weight,mode=text}
    \renewrobustcmd{\bfseries}{\fontseries{b}\selectfont}
    \renewrobustcmd{\boldmath}{}
    \newrobustcmd{\B}{\bfseries}
    \caption{DERs (\%) on LibriCSS and AMI Mix-Headset.}\label{tbl:diarization_libricss}
    \setlength{\tabcolsep}{3pt}
    \resizebox{\linewidth}{!}{%
    \begin{tabular}{@{}l*{3}{S[table-format=1.2]}*{4}{S[table-format=2.2]}@{}}
        \toprule
        &\multicolumn{6}{c}{LibriCSS}&{AMI}\\\cmidrule(l{\tabcolsep}r{\tabcolsep}){2-7}\cmidrule(l{\tabcolsep}){8-8}
        Method&{0L}&{0S}&{OV10}&{OV20}&{OV30}&{OV40}&{Mix-Headset}\\\midrule
        SC \cite{park2020auto} & 1.36 & 0.32 & 7.45 & 14.28 & 19.78 & 23.36&18.95\\
        SC-OL \cite{raj2021multiclass} & \B 1.28 & \B 0.18 & \B 6.85 & 11.84 & 16.63 & 18.06&17.04\\
        SC + Proposed & 1.29 & 0.75 & 7.10 & \B 8.31 & \B 11.87 & \B 12.71&\B 16.93\\
        \bottomrule
    \end{tabular}%
    }
\end{table}

\autoref{tbl:diarization_libricss} shows the DERs on the LibriCSS dataset and AMI Mix-Headset corpus.
Focusing on the results on LibriCSS, the proposed method showed slightly degraded DERs when the overlap ratio was small but significantly improved DERs when the overlap ratio was high relative to SC-OL.
With the improvements in the case of high overlap ratios, it can be said that the proposed speaker embedding extractor enables us to extract two speaker embeddings from one interval, and thus, we can perform overlap-aware diarization by simply clustering them.
The proposed method also outperformed the baselines on the AMI Mix-Headset corpus, further supporting its effectiveness.
The slight degradations in DER with small overlap ratios are considered to be due to a marginal degradation in single-speaker embeddings (cf. \svss{} in \autoref{tbl:main_results}), indicating room for improvement in the proposed method.

\section{Conclusion}
In this paper, we proposed a method for extracting speaker embeddings from multi-speaker audio.
The proposed method enabled the extraction of speaker embedding for each speaker, even from fully overlapped speech, by recursively calculating the attention weights for pooling.
We also proposed a method for simultaneously estimating the number of speakers based on the calculated attention weights.
Experimental results showed that the proposed method offers improvements in both speaker verification and diarization performance.
Future work will include the combination of the proposed method with end-to-end diarization framework.

\clearpage
\bibliographystyle{IEEEbib}
\bibliography{mybib}

\end{document}